\documentclass[aps,pre,twocolumn,showpacs,superscriptaddress,groupedaddress]{revtex4-1}
\pdfoutput=1
\usepackage{hyperref}
\usepackage{epsfig}
\usepackage{amsmath}
\usepackage{graphicx}
\usepackage{wrapfig}
\usepackage{dcolumn}
\usepackage{bm}
\usepackage{amssymb}
\usepackage{titlesec}
\usepackage{mathtools}
\usepackage{relsize}
\usepackage{cleveref}
\usepackage[usenames,dvipsnames]{color}

\begin{document}
\title{Unstable oscillations and bistability in delay-coupled swarms}
\author{Jason Hindes$^{1}$, Victoria Edwards$^{2}$, Sayomi Kamimoto$^{3}$, Ioana Triandaf$^{1}$, and Ira B. Schwartz$^{1}$}
\affiliation{$^{1}$U.S. Naval Research Laboratory, Code 6792, Plasma Physics Division, Washington, DC 20375, USA}
\affiliation{$^{2}$U.S. Naval Research Laboratory, Code 5514, Navy Center for Applied Research in Artificial Intelligence, Washington, DC 20375, USA}
\affiliation{$^{3}$Department of Mathematics, George Mason University, Fairfax Virginia, 22030, USA}

\begin{abstract}
It is known from both theory and experiments that introducing time delays into the communication network of
mobile-agent swarms produces coherent rotational patterns. 
Often such spatio-temporal rotations can be bistable 
with other swarming patterns, such as milling and flocking. Yet, most known
bifurcation results related to delay-coupled swarms rely on inaccurate
mean-field techniques. As a consequence, the utility of applying macroscopic theory as a guide for predicting
and controlling swarms of mobile robots has been limited. To overcome this limitation, we perform an exact stability analysis of two primary swarming patterns in a general model with time-delayed interactions. By correctly identifying the relevant spatio-temporal modes that determine stability in the presence of time delay, we are able to accurately predict bistability and unstable oscillations in large swarm simulations-- laying the groundwork for comparisons to robotics experiments.   
\end{abstract}
\maketitle

\section{\label{sec:Intro}INTRODUCTION}
In nature, swarms consist of individual agents with limited dynamics and simple rules, which interact, sense, collaborate and actuate to produce emergent spatio-temporal patterns. Examples include schools of fish\cite{Couzin2013, Calovi2014, Cavagna2015}, flocks of starlings\cite{Leonard2013, Ballerini08} and jackdaws\cite{Ouellette2019}, colonies of bees\cite{Li_Sayed_2012}, ants\cite{Theraulaz2002}, locusts\cite{Topaz2012}, and bacteria\cite{Polezhaev}, as well as crowds of people\cite{Rio_Warren_2014}. Given the many examples across a wide range of space and time scales, significant progress has been made in understanding swarming by studying simple dynamical models with general properties\cite{Vicsek,Marchetti,Aldana}. 

Deriving inspiration from nature, embodied artificial swarm systems have been created to mimic emergent pattern formation-- with the ultimate goal of designing robotic swarms that can perform complex tasks {\it autonomously}\cite{Brambilla2013, Bayindir16,Bandyopadhyay, Wu}. Recently robotic swarms have been used experimentally for applications such as mapping\cite{Ramachandran2018}, leader-following\cite{MORGAN2005,Wiech2018}, and density control\cite{Li17}. To achieve swarming behavior, often, robots are controlled based on models, where swarm properties can be predicted exactly\cite{Tanner07, Gazi05, Jadbabaie03, Viraghn14, Desai01}. Such approaches rely on strict assumptions to guarantee behavior. Any uncharacterized dynamics can cause patterns to be lost or changed. This is particularly the case for robotic swarms that move in uncertain environments and must satisfy realistic communication constraints.

In particular in both robotic and biological swarms, there is often a delay between the time information is perceived and the reaction time of agents. 
Such delays have been measured in swarms of bats\cite{Luca_bats}, birds\cite{Nagy_pigeons}, fish\cite{JL_fish}, and crowds of people\cite{JF_people}. Delays naturally occur in robotic swarms communicating over wireless networks, due to low bandwidth\cite{Komareji2018} and multi-hop communication\cite{Oliveira15}. In general, time-delays in swarms result in multi-stability of rotational patterns in space, and the possibility of switching between patterns\cite{Romero2012, Szwaykowska2016, Edwards2019,Wells15, Hartle17, Ansmann16, Hindes18, Szwaykowska2018, Kularatne19}. Though observed in simulations and experiments, swarm bistability due to time-delay has lacked an accurate quantitative description, which we provide in this work. 

Consider a system of mobile agents, or swarmers, moving under the influence of three forces: self-propulsion, friction, and mutual attraction.  
In the absence of attraction, each swarmer tends to a fixed speed, which balances propulsion and friction but has no preferred direction. The agents 
are assumed to communicate through a network with time delays. Namely, each agent is attracted to where its neighbors were at some moment in the past. A simple model 
which captures the basic physics is
\begin{align}
m\ddot{\bold{r}}_{i}= \big[\alpha -\beta|\dot{\bold{r}}_{i}|^{2}\big]\dot{\bold{r}}_{i}+\frac{a}{N-1}\sum_{j\neq i}[\bold{r}_{j}(t-\tau)-\bold{r}_{i}]+\boldsymbol{\xi}_{i}(t). 
\label{eq:BasicPhysics}
\end{align} 
where $m$ is the mass of each agent, $\alpha$ is a self-propulsion constant,
$\beta$ is a friction constant, $a$ is a swarm coupling constant, $\tau$ is a
characteristic time delay, $N$ is the number of agents, $\bold{r}_{i}$  is the
position-vector for the $i$th agent in two spatial dimensions, and
$\boldsymbol{\xi}_{i}(t)$ is a small noise
source\cite{Levine,Erdmann,DOrsagna,F1,Romero2012}. Note: in this work we consider the
simple case of spring interaction forces and global communication topology for
illustration and ease of analysis; however, these assumptions can be relaxed with predictable effects on the dynamics\cite{Chuang,Bernoff,J1,Szwaykowska2016}. 

Recently, Eq.(\ref{eq:BasicPhysics}) has been implemented in experiments with several robotics platforms including autonomous cars, boats, and (large and small) quad-rotors\cite{Szwaykowska2016,Edwards2019}. In all such experiments, a number of robots move in space, and each robot's position is measured by a motion-capture system. This data is fed into a simulator which in addition to the robots' positions, maintains the position of a larger number of simulated agents. New velocity commands are sent over a wireless network to each robot according to Eq.(\ref{eq:BasicPhysics})\cite{Szwaykowska2016,Edwards2019}-- with the addition of small repulsive forces to keep the robots from colliding.

\section{\label{sec:UnstableModes}SWARMING PATTERNS AND STABILITY}
From generic initial conditions a swarm described by Eq.(\ref{eq:BasicPhysics}) typically tends to one of two stable spatio-temporal patterns: a ring (milling) state, or a rotating state -- depending on initial conditions and parameters\cite{F1}. The two patterns can be seen in Fig.\ref{fig:StabilityDiagram} (b). Note that the snapshots in time are drawn from simulations of Eq.(\ref{eq:BasicPhysics}) with small noise. The emergence and stability of the ring and rotating patterns are often qualitatively described using mean-field approximations, in which the motions of agents relative to the swarm's center-of-mass are neglected\cite{Romero2012,Lindley2013}. Though useful, such coarse descriptions do not capture bistability and noise-induced switching, let alone the more complex motions observed in swarming experiments\cite{Edwards2019,Szwaykowska2016}. What's more, higher-order approximation techniques predict bistability qualitatively, but still suffer from quantitative inaccuracy, and are difficult to analyze\cite{Romero2014}. Hence, an analyzable and  accurate description of stability is needed, especially for robotics experiments which use Eq.(\ref{eq:BasicPhysics}) (and its generalizations) as a basic autonomy-controller. In support of such experimental efforts, below, we analyze the linear stability of the ring and rotating states exactly for large $N$ and compare our predictions to numerical simulations.

\subsection{\label{subsec:Ring}Ring State}
First, it is useful to transform Eq.(\ref{eq:BasicPhysics}) into polar coordinates where the ring and rotating states can be naturally represented as fixed-point solutions in appropriately chosen rotating reference frames. Introducing the coordinate transformations $\bold{r}_{i}\!\equiv\!\left<r_{i}\cos{\!(\phi_{i})}\;,r_{i}\sin{\!(\phi_{i})}\right>$, substituting into Eq.(\ref{eq:BasicPhysics}), and neglecting noise, we obtain:
\begin{align}
\label{eq:Polar1}
mr_{i}\ddot{\phi_{i}}=&\big[\alpha-\beta\big(r_{i}^{2}\dot{\phi}_{i}^{2}+\dot{r}_{i}^{2}\big)\big]r_{i}\dot{\phi}_{i}-2m\dot{r}_{i}\dot{\phi}_{i}\\ \nonumber 
                                  &+\frac{a}{N-1}\!\sum_{j\neq i}r_{j}(t-\tau)\sin\!\big(\phi_{j}(t-\tau)-\phi_{i}\big),\\
\label{eq:Polar2}
m\ddot{r}_{i}=&\big[\alpha-\beta\big(r_{i}^{2}\dot{\phi}_{i}^{2}+\dot{r}_{i}^{2}\big)\big]\dot{r}_{i} +mr_{i}\dot{\phi}_{i}^{2}\\ \nonumber 
                                  &+\frac{a}{N-1}\!\sum_{j\neq i}\big[r_{j}(t-\tau)\cos\!\big(\phi_{j}(t-\tau)-\phi_{i}\big)-r_{i}\big].                                   
\end{align}

\begin{figure}[h]
\vspace{0.65cm}
\includegraphics[scale=0.3265]{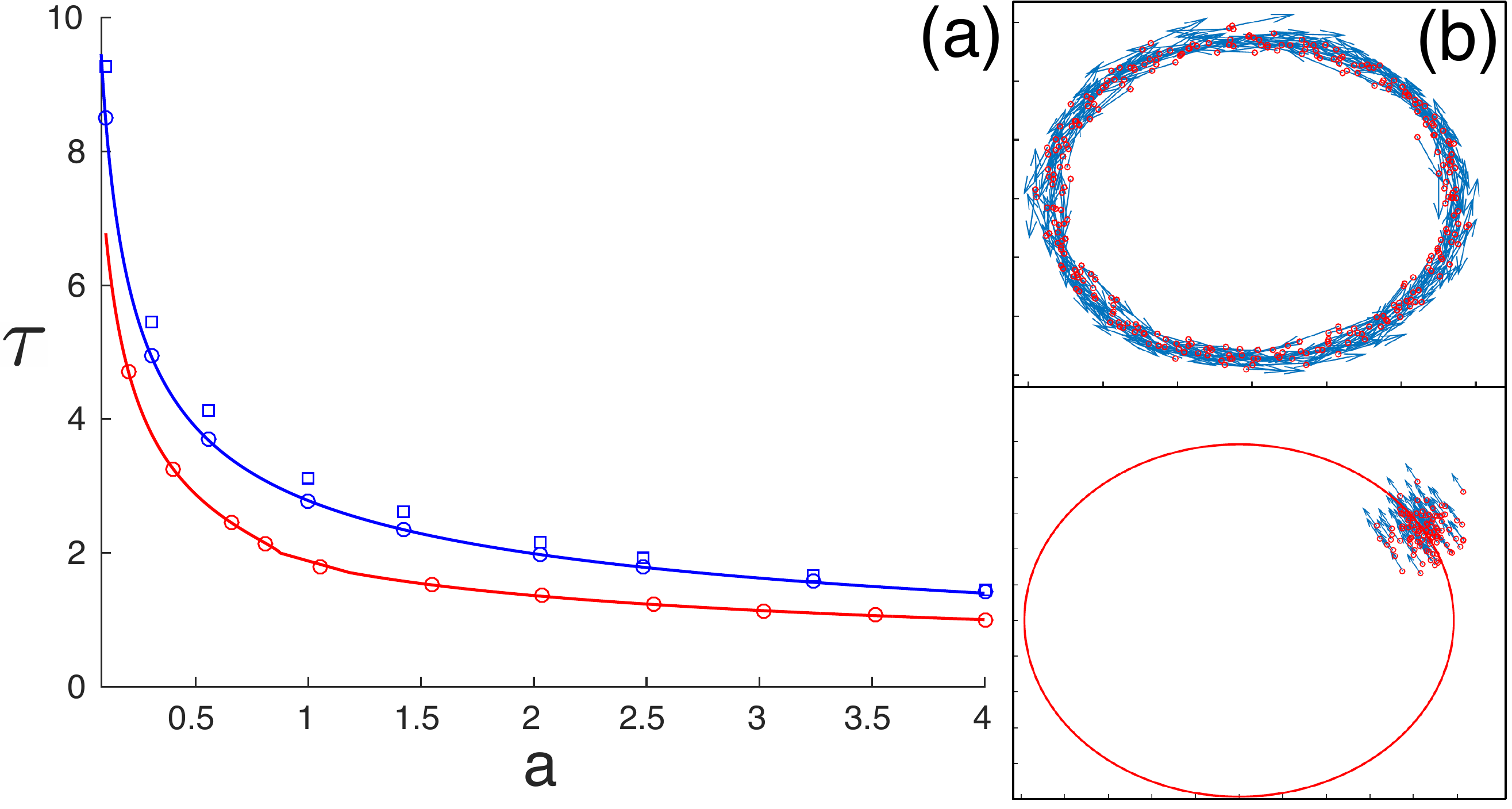}
\caption{Stability diagram for delay-coupled swarms. (a) The blue curve denotes a Hopf bifurcation for the ring state (b, upper). The red curve denotes a combined line for saddle-node and double-Hopf bifurcations for the rotating state (b, lower). Points denote simulation-determined stability changes for: a ring state with all agents rotating in the same direction (blue circles), a ring state with half the agents rotating in the opposite direction (blue squares), and a rotating state (red circles). (b) Snapshots for both states in the x-y plane ($a\!=\!1,\; \tau\!=\!2.6$,\; N=100). Agents are drawn with red circles. Velocities are drawn with blue arrows.}
\label{fig:StabilityDiagram}
\end{figure}

When $N\!\gg\!1$\cite{Note2}, ring-state formations correspond to solutions of Eqs.(\ref{eq:Polar1}-\ref{eq:Polar2}) where radii and phase-velocities are constant\cite{Romero2012}, and phases are uniformly distributed, e.g.,    
\begin{align}
\label{eq:Ring}
r_{j}(t)=\sqrt{\frac{m\alpha}{\beta a}}, \;\;\;\;\;\; \phi_{j}(t)=\frac{2\pi(j-1)}{N}+\sqrt{\frac{a}{m}}t.                                
\end{align}
This is easy to check by direct substitution. In general, many related ring states also exist, i.e, where some number of agents have the opposite angular velocity, $-\sqrt{a/m}$, and are distributed uniformly around a concentric ring. In our stability analysis below, we focus on the case where all agents rotate in the same direction for three reasons: this case persists when small repulsive forces are added (as in robotics experiments\cite{Szwaykowska2016,Edwards2019}), the stability of any given ring pattern has only a weak dependence on the number of nodes rotating in each direction (as demonstrated with simulations), and analytical tractability.
 
To determine the local stability of the ring state we need to understand how small perturbations to Eq.(\ref{eq:Ring}) grow (or decay) in time. Our first step is to substitute a general perturbation, $r_{j}(t)\!=\!\sqrt{m\alpha/\beta a}+\!B_{j}(t)$ and $\phi_{j}(t)\!=\!2\pi(j-1)/N\!+\!\sqrt{a/m}\;t +\!A_{j}(t)$, into Eqs.(\ref{eq:Polar1}-\ref{eq:Polar2}) and collect terms to first order in $A_{j}(t)$ and $B_{j}(t)$ (assuming $A_{j}\;, B_{j} \ll1\;\forall j$). The result is the following linear system of delay-differential equations for $N\gg1$ {\it with constant coefficients} -- the latter property is a consequence of our transformation into the proper coordinate system and is what allows for an analytical treatment:
\begin{widetext} 
\begin{align}
\label{eq:LinRing1}
&\sqrt{\frac{m\alpha}{\beta a}}\big[m\ddot{A}_{i}+2\alpha\dot{A}_{i}\big]+2\sqrt{\!\frac{a}{m}}\big[m\dot{B}_{i}+\alpha B_{i}\big]=\frac{a}{N}\!\sum_{j}\!\Big[B_{j}^{\tau}\sin\!\Big(\!\tfrac{2\pi (j-i)}{N}-\!\!\sqrt{\!\tfrac{a}{m}}\tau\!\Big)+\!(A_{j}^{\tau}-\!A_{i})\sqrt{\frac{m\alpha}{\beta a}}\cos\!\Big(\!\tfrac{2\pi (j-i)}{N}-\!\!\sqrt{\!\tfrac{a}{m}}\tau\!\Big)\Big],\\
\label{eq:LinRing2}
&m\ddot{B}_{i}-2m\sqrt{\frac{\alpha}{\beta}}\dot{A}_{i}=\frac{a}{N}\!\sum_{j}\!\Big[B_{j}^{\tau}\cos\!\Big(\!\tfrac{2\pi (j-i)}{N}-\!\!\sqrt{\!\tfrac{a}{m}}\tau\!\Big)-\!(A_{j}^{\tau}-\!A_{i})\sqrt{\frac{m\alpha}{\beta a}}\sin\!\Big(\!\tfrac{2\pi (j-i)}{N}\!-\!\!\sqrt{\!\tfrac{a}{m}}\tau\!\Big)\Big]                             
\end{align}
\end{widetext}
where $A_{j}^{\tau}\!\equiv\!A_{j}(t-\tau)$ and $B_{j}^{\tau}\!\equiv\!B_{j}(t-\tau)$.  

Given the periodicity implied by the equally-spaced phase variables in Eq.(\ref{eq:Ring}), it is natural to look for eigen-solutions of Eqs.(\ref{eq:LinRing1}-\ref{eq:LinRing2}) in terms of the discrete Fourier transforms of $A_{j}(t)$ and $B_{j}(t)$. In fact, by inspection  we can see that only the first harmonic survives the summations on the right-hand sides of Eqs.(\ref{eq:LinRing1}-\ref{eq:LinRing2}), because of the sine and cosine terms, and hence we look for particular solutions: $A_{j}(t)=A\exp\{\lambda t -2\pi i(j-1)/N\}$ and $B_{j}(t)=B\exp\{\lambda t -2\pi i(j-1)/N\}$. Substitution and a fair bit of algebra gives the following transcendental equation for the stability exponent, $\lambda$, of the ring state:
\begin{align}
\label{eq:RingStability}
&\frac{m\lambda^{2}+\!2\alpha\lambda-\!\frac{a}{2}e^{\!-\tau\!\big[\lambda+i\sqrt{\!\frac{a}{m}}\;\big]}}{2m\sqrt{\!\frac{a}{m}}\lambda-\!\frac{a}{2i}e^{\!-\tau\!\big[\lambda+i\sqrt{\!\frac{a}{m}}\;\big]}} \;\;\;\;\;\;\;\;\;\;+ \nonumber 
\\
&\frac{2\sqrt{\!\frac{a}{m}}\big[m\lambda+\alpha\big]-\!\frac{a}{2i}e^{\!-\tau\!\big[\lambda+i\sqrt{\!\frac{a}{m}}\;\big]}}{m\lambda^{2}-\!\frac{a}{2}e^{\!-\tau\!\big[\lambda+i\sqrt{\!\frac{a}{m}}\;\big]}} \;\;\;\;\;\;\;=0.                               
\end{align}

In general, the ring state will be linearly stable if there are no solutions to Eq.(\ref{eq:RingStability}) with $Re[\lambda]\!>\!0$. In fact, varying $a$ and $\tau$ while fixing the other parameters, we discover a Hopf bifurcation, generically, at which $\lambda\!=\!\pm i\omega_{c}$ \cite{Kuznetsov1}. An example Hopf line is shown in Fig.\ref{fig:StabilityDiagram}(a) in blue for $m\!=\!\alpha\!=\!\beta\!=1.$ Based on our analysis, we expect the ring state to be locally stable below the blue line and unstable above it. For comparison, the blue circles in Fig.\ref{fig:StabilityDiagram}(a) denote simulation-determined transition points: the smallest $\tau(a)$ for which a swarm of 600 agents, initially prepared in a ring state with a small random perturbation (i.e., independent and uniformly distributed $A_{j}$ and $B_{j}$ over $[-10^{-5},10^{-5}]$), returns to a ring configuration after an integration time of $t=20000$. Numerical predictions from Eq.(\ref{eq:RingStability}) show excellent agreement with these simulation results. Similarly determined transition points for a ring formation in which half the agents rotate in one direction, and half rotate in the opposite direction, are shown with blue squares. We can see that the ring's Hopf-transition line still gives a good approximation for this more general case, especially for larger values of $a$.
\begin{figure}[h]
\vspace{0.27cm}
\includegraphics[scale=0.28]{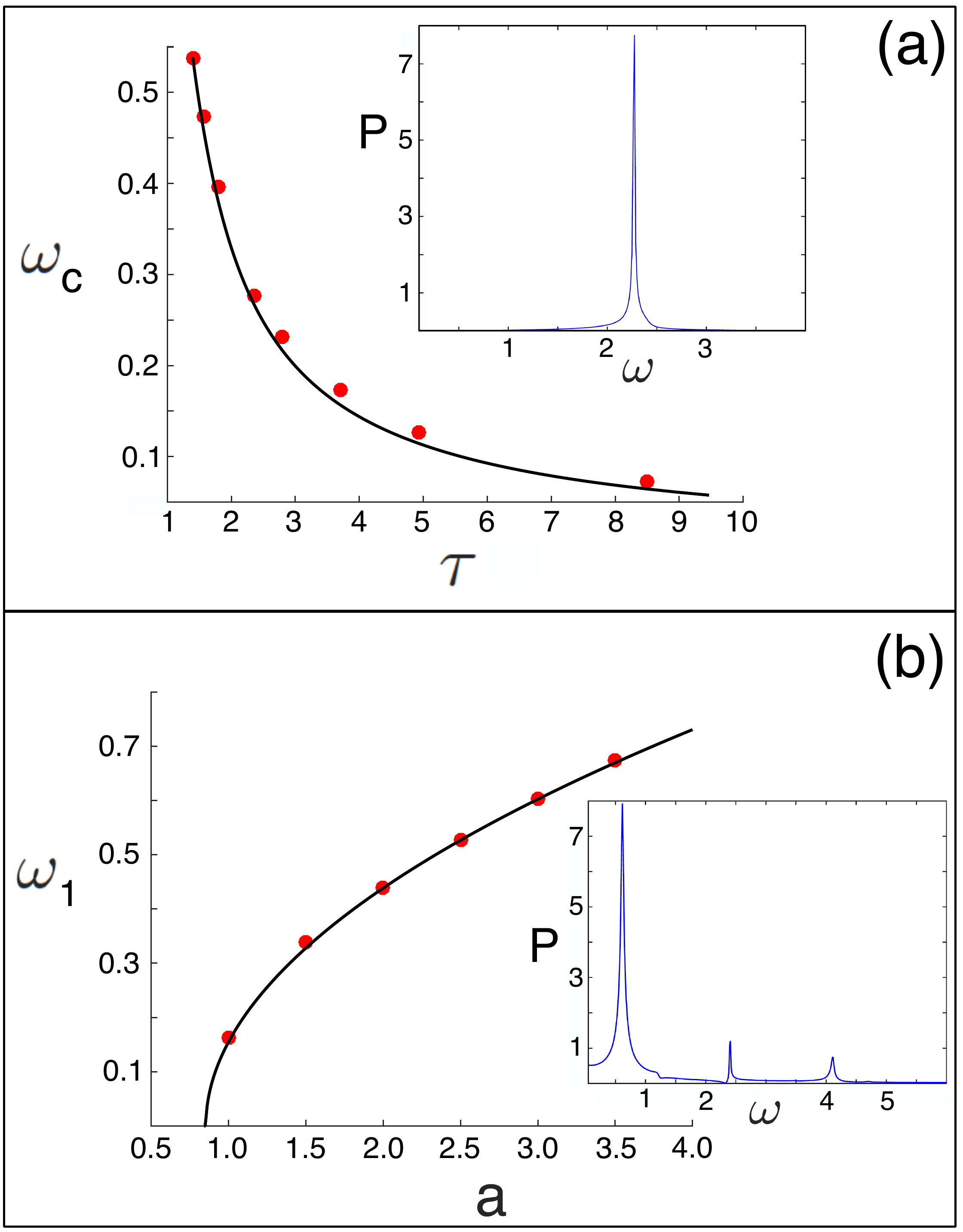}
\caption{Frequency of unstable modes near bifurcation. (a) unstable frequency for the ring state at the Hopf bifurcation (black line) determined from the power spectrum of the swarm's center-of-mass (red circles). (b) unstable frequency for the rotating state at the double-Hopf bifurcation (black line) determined from the power spectrum for a single agent. Inlet panels show example spectra for both states.}
\label{fig:UnstableMode}
\end{figure}
 
In addition to the transition points, we can check the frequency of oscillations around the ring state, implied by the existence of an unstable mode for $\tau(a)$ slightly above the Hopf bifurcation. First we perform a simulation initially prepared in the ring state with a small perturbation (as described in the preceding paragraph), and compute the peak frequency, $\omega^{*}$, in the Fourier spectrum of the swarm's center-of-mass, $\bold{R}(t)\!\equiv\!\sum_{j}\bold{r}_{j}/N$. An example is shown in the inlet panel of Fig.\ref{fig:UnstableMode}(a) for ($a\!=\!3.243$, $\tau\!=\!1.565$); the symbol $\text{P}$ denotes the absolute value of the Fourier transform. Second, we plot $\omega_{c}\!=\!\omega^{*}\!-\!\sqrt{a/m}$ and compare to predictions from solutions of Eq.(\ref{eq:RingStability}) with $\lambda\!=\!\pm i\omega_{c}\!\neq\!0$ for a range of time delays. The comparison is shown in Fig.\ref{fig:UnstableMode}(a) with excellent agreement.     

\subsection{\label{sec:Rotating}Rotating State}
Next, we perform a similar stability analysis for the rotating state, which has a different bifurcation structure and unstable modes. Unlike the ring state, the rotating state entails a collapse of the swarm on to the center of mass (in the noiseless limit). In polar coordinates, the agents satisfy $r_{j}(t)\!=\!R$ and $\phi_{j}(t)\!=\!\Omega t$ \cite{Romero2012}, where 
\begin{align}
\label{eq:Rotating1}
&0=m\Omega^{2}-\!a\big[1\!-\cos{\Omega\tau}\big],\\  
\label{eq:Rotating2}
&R=\frac{1}{|\Omega|}\sqrt{\frac{\alpha-a\sin(\Omega\tau)/\Omega}{\beta}}.                          
\end{align}
In order to determine the local stability of the rotating state we substitute
$r_{j}(t)\!=\!R+\!B_{j}\exp\{\lambda t\}$ and $\phi_{j}(t)\!=\!\Omega\;t
+A_{j}\exp\{\lambda t\}$, into Eqs.(\ref{eq:Polar1}-\ref{eq:Polar2}) and,
again, collect terms to first order in $A_{j}$ and $B_{j}$ (assuming $A_{j}\;,
B_{j} \ll1\;\forall j$). The result is another linear system of equations with constant
coefficients. After some algebra, we obtain for $N\!\gg\!1$:
\begin{widetext} 
\begin{align}
\label{eq:LinRotating1}
R\big[m\lambda^{2}-\!\lambda(\alpha-\!3\beta R^{2}\Omega^{2})+\!a\cos(\Omega\tau)\big]A_{i}+\Omega\big[2m\lambda -\!\alpha+\!3\beta R^{2}\Omega^{2}\big]B_{i}=\;&\frac{ae^{-\!\lambda\tau}}{N}\sum_{j}\!\Big[R\cos(\Omega\tau)A_{j}
-\sin(\Omega\tau)B_{j}\!\Big], \\
\label{eq:LinRotating2}
\big[aR\sin(\Omega\tau)-\!2mR\Omega\lambda\big]A_{i}+\big[m\lambda^{2}-m\Omega^{2}-\!\lambda(\alpha-\!\beta R^{2}\Omega^{2})+a]B_{i}=\; &\frac{ae^{-\!\lambda\tau}}{N}\sum_{j}\!\Big[R\sin(\Omega\tau)A_{j}+\cos(\Omega\tau)B_{j}\!\Big].                                
\end{align}
 \end{widetext} 
 
There are two primary categories of solutions to Eqs.(\ref{eq:LinRotating1}-\ref{eq:LinRotating2}). The first is $A_{i}\!=\!A$ and $B_{i}\!=\!B$, which we call the homogeneous modes. Because all agents move together (equal to the the center-of-mass motion) the stability entailed by the homogeneous modes should match the mean-field approximation mentioned above and analyzed in\cite{Romero2012}. Because the mean-field is known to be quantitatively inaccurate for capturing stability\cite{Romero2014,Edwards2019}, we focus on the second set of solutions: $\sum_{j}A_{j}/N\!=0$ and $\sum_{j}B_{j}/N\!=0$. The stability-exponents, $\lambda$, for these modes satisfy     
\begin{align}
\label{eq:LocalRotating}
&\frac{2m\Omega\lambda-a\sin(\Omega\tau)}{m\lambda^{2}-\lambda(\alpha-3\beta R^{2}\Omega^{2}) +a\cos(\Omega\tau)} \;\;- \nonumber \\ 
&\frac{m\lambda^{2}-m\Omega^{2}-\lambda(\alpha-\beta R^{2}\Omega^{2})+a}{\Omega\big[\alpha-3\beta R^{2}\Omega^{2}-2m\lambda\big]}\;\;\;\;\;\;\;\;=0,                                
\end{align}
which has four complex solutions.
 
In general, the rotating state will be linearly stable if there are no solutions to Eq.(\ref{eq:LocalRotating}) with $Re[\lambda]\!>\!0$. In practice, we find that changing $a$ and $\tau$ while keeping all other parameters fixed, produces saddle-node, Hopf and double-Hopf bifurcations\cite{Note1,Kuznetsov1,J2}. In the former case, a single real eigenvalue approaches zero, or  
\begin{align}
\label{eq:SNRotating}
&\tan(\Omega\tau)=\frac{m\Omega^{2}-a}{\Omega(\alpha-3\beta R^{2}\Omega^{2})}.                              
\end{align}
Equation (\ref{eq:SNRotating}) gives the stability-line for the rotating state with small $a$ and large $\tau$. For large $a$ and small $\tau$, the stability changes through a double-Hopf bifurcation where two frequencies become unstable simultaneously, $\lambda\!=\!\pm i\omega_{1},\;\pm i\omega_{2}\!\neq\!0$. Fig.\ref{fig:StabilityDiagram}(a) shows the predicted composite stability-curve for the rotating state, combining both bifurcations. Plotted is the maximum $\tau$, for fixed $a$, where $Re[\lambda]\!>\!0$. Above the red line the rotating state is expected to be locally stable, and below it, unstable (see Sec.\ref{sec:App} for an enlarged view of the stability diagram).  

As with the ring state, we compare our stability predictions to simulations, and determine the smallest value of $\tau$ for which a swarm of $N\!=\!600$ agents, initially prepared in the rotating state with a small, random perturbation, returns to a rotating state after a time of $t=20000$. These points are shown with red circles in Fig.\ref{fig:StabilityDiagram}(a) for several values of coupling. Again, we find excellent agreement with predictions. Another consequence of our analysis is the clear quantitative prediction of swarm bistability (between the red and blue curves) and noise-induced switching between ring and rotating patterns, which can now be precisely tested in experiments\cite{Szwaykowska2016,Edwards2019,Szwaykowska2018}.

Lastly, just as with the ring state, we can compare the frequency of oscillations around the rotating state for $\tau(a)$ slightly below the double-Hopf bifurcation values, where we expect weak instability of modes orthogonal to the center-of-mass-motion. First we perform a simulation initially prepared in the rotating state with a small perturbation, and compute the peak frequency, $\omega^{*}$, in the Fourier spectrum of $r_{j}-R$, where $j$ is a randomly selected agent. An example is shown in the inlet panel of Fig.\ref{fig:UnstableMode}(b) for ($a\!=\!3.5$, $\tau\!=\!1.059$). This peak frequency is compared to predictions from numerical solutions of Eq.(\ref{eq:LocalRotating}) with $\lambda\!=\!\pm i\omega_{1},\;\pm i\omega_{2}\!\neq\!0$ for a range of coupling strengths. In Fig.\ref{fig:UnstableMode}(b) the smaller of the two frequencies, $\omega_{1}$ is plotted along with $\omega^{*}$ -- showing excellent agreement. Note that in this comparison, we do not subtract off the rotating state's frequency, $\Omega$, since $r_{j}$ does not oscillate in the rotating state but is equal to $R$.\\     

\section{CONCLUSION}
In this work we studied the stability of ring and rotational patterns in a general swarming model with time-delayed interactions. We found that ring states change stability through Hopf bifurcations, where spatially periodic modes sustain oscillations in time. On the other hand, rotating states undergo saddle-node, Hopf, and double-Hopf bifurcations, where modes with orthogonal dynamics to the center-of-mass-motion change stability. For both states, the unstable oscillations correspond to dynamics not captured by standard mean-field approximations. Our results were verified in detail with large-agent simulations. Future work will extend our analysis to include the effects of repulsive forces, noise, and incomplete (and dynamic) communication topology -- all of which are necessary for parametrically controlling real swarms of mobile robots.  

JH, IT, and IBS were supported by the U.S. Naval Research Laboratory funding (N0001419WX00055) and the Office of Naval Research (N0001419WX01166) and (N0001419WX01322). TE was supported through the U.S Naval Research Laboratory Karles Fellowship. SK was supported through the GMU Provost PhD award as part of the Industrial Immersion Program\pagebreak

\begin{figure}[t]
\includegraphics[scale=0.28]{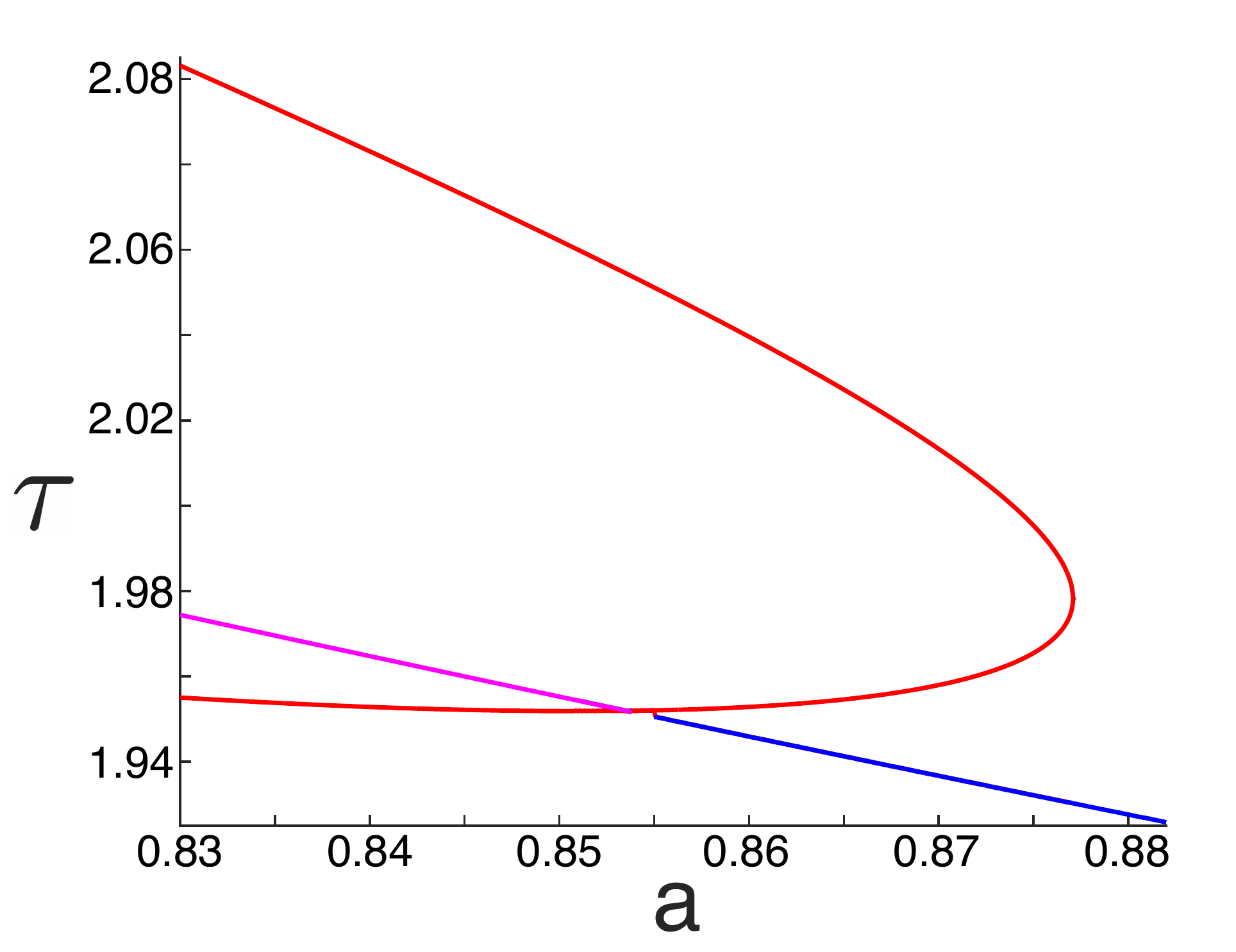}
\caption{Bifurcation curves for the rotating state near the intersection of saddle-node (red), Hopf (magenta), and double-Hopf (blue) bifurcations.}
\label{fig:Zoom}
\end{figure}

\appendix{\centerline{\bf IV. APPENDIX}\label{sec:App}}
Close inspection of Fig.\ref{fig:StabilityDiagram}(a), shows that there is a
small, apparent discontinuity in the stability line for the rotating
state. This apparent discontinuity is a consequence of several bifurcation
curves intersecting in a small region in the $(a,\tau)$ plane for
$m\!=\!\alpha\!=\!\beta\!=\!1$. For completeness, we show an enlarged version
of the stability diagram for the rotating state in Fig.\ref{fig:Zoom}. The
rotating state is linearly stable in the region bounded to the left by the red
(saddle-node) and blue (double-Hopf) bifurcations.

\bibliography{DelayBistabilityPRE1}

\begin{thebibliography}{55}%
\makeatletter
\providecommand \@ifxundefined [1]{%
 \@ifx{#1\undefined}
}%
\providecommand \@ifnum [1]{%
 \ifnum #1\expandafter \@firstoftwo
 \else \expandafter \@secondoftwo
 \fi
}%
\providecommand \@ifx [1]{%
 \ifx #1\expandafter \@firstoftwo
 \else \expandafter \@secondoftwo
 \fi
}%
\providecommand \natexlab [1]{#1}%
\providecommand \enquote  [1]{``#1''}%
\providecommand \bibnamefont  [1]{#1}%
\providecommand \bibfnamefont [1]{#1}%
\providecommand \citenamefont [1]{#1}%
\providecommand \href@noop [0]{\@secondoftwo}%
\providecommand \href [0]{\begingroup \@sanitize@url \@href}%
\providecommand \@href[1]{\@@startlink{#1}\@@href}%
\providecommand \@@href[1]{\endgroup#1\@@endlink}%
\providecommand \@sanitize@url [0]{\catcode `\\12\catcode `\$12\catcode
  `\&12\catcode `\#12\catcode `\^12\catcode `\_12\catcode `\%12\relax}%
\providecommand \@@startlink[1]{}%
\providecommand \@@endlink[0]{}%
\providecommand \url  [0]{\begingroup\@sanitize@url \@url }%
\providecommand \@url [1]{\endgroup\@href {#1}{\urlprefix }}%
\providecommand \urlprefix  [0]{URL }%
\providecommand \Eprint [0]{\href }%
\providecommand \doibase [0]{http://dx.doi.org/}%
\providecommand \selectlanguage [0]{\@gobble}%
\providecommand \bibinfo  [0]{\@secondoftwo}%
\providecommand \bibfield  [0]{\@secondoftwo}%
\providecommand \translation [1]{[#1]}%
\providecommand \BibitemOpen [0]{}%
\providecommand \bibitemStop [0]{}%
\providecommand \bibitemNoStop [0]{.\EOS\space}%
\providecommand \EOS [0]{\spacefactor3000\relax}%
\providecommand \BibitemShut  [1]{\csname bibitem#1\endcsname}%
\let\auto@bib@innerbib\@empty
\bibitem [{\citenamefont {Tunstrøm}\ \emph {et~al.}(2013)\citenamefont
  {Tunstrøm}, \citenamefont {Katz}, \citenamefont {Ioannou}, \citenamefont
  {Huepe}, \citenamefont {Lutz},\ and\ \citenamefont {Couzin}}]{Couzin2013}%
  \BibitemOpen
  \bibfield  {author} {\bibinfo {author} {\bibfnamefont {K.}~\bibnamefont
  {Tunstrøm}}, \bibinfo {author} {\bibfnamefont {Y.}~\bibnamefont {Katz}},
  \bibinfo {author} {\bibfnamefont {C.~C.}\ \bibnamefont {Ioannou}}, \bibinfo
  {author} {\bibfnamefont {C.}~\bibnamefont {Huepe}}, \bibinfo {author}
  {\bibfnamefont {M.~J.}\ \bibnamefont {Lutz}}, \ and\ \bibinfo {author}
  {\bibfnamefont {I.~D.}\ \bibnamefont {Couzin}},\ }\href {\doibase
  10.1371/journal.pcbi.1002915} {\bibfield  {journal} {\bibinfo  {journal}
  {PLoS. Comput. Biol.}\ }\textbf {\bibinfo {volume} {9}},\ \bibinfo {pages}
  {1} (\bibinfo {year} {2013})}\BibitemShut {NoStop}%
\bibitem [{\citenamefont {Calovi}\ \emph {et~al.}(2014)\citenamefont {Calovi},
  \citenamefont {Lopez}, \citenamefont {Ngo}, \citenamefont {Sire},
  \citenamefont {Chat{\'{e}}},\ and\ \citenamefont {Theraulaz}}]{Calovi2014}%
  \BibitemOpen
  \bibfield  {author} {\bibinfo {author} {\bibfnamefont {D.~S.}\ \bibnamefont
  {Calovi}}, \bibinfo {author} {\bibfnamefont {U.}~\bibnamefont {Lopez}},
  \bibinfo {author} {\bibfnamefont {S.}~\bibnamefont {Ngo}}, \bibinfo {author}
  {\bibfnamefont {C.}~\bibnamefont {Sire}}, \bibinfo {author} {\bibfnamefont
  {H.}~\bibnamefont {Chat{\'{e}}}}, \ and\ \bibinfo {author} {\bibfnamefont
  {G.}~\bibnamefont {Theraulaz}},\ }\href {\doibase
  10.1088/1367-2630/16/1/015026} {\bibfield  {journal} {\bibinfo  {journal}
  {New Journal of Physics}\ }\textbf {\bibinfo {volume} {16}},\ \bibinfo
  {pages} {015026} (\bibinfo {year} {2014})}\BibitemShut {NoStop}%
\bibitem [{\citenamefont {Cavagna}\ \emph {et~al.}(2015)\citenamefont
  {Cavagna}, \citenamefont {Del~Castello}, \citenamefont {Giardina},
  \citenamefont {Grigera}, \citenamefont {Jelic}, \citenamefont {Melillo},
  \citenamefont {Mora}, \citenamefont {Parisi}, \citenamefont {Silvestri},
  \citenamefont {Viale},\ and\ \citenamefont {Walczak}}]{Cavagna2015}%
  \BibitemOpen
  \bibfield  {author} {\bibinfo {author} {\bibfnamefont {A.}~\bibnamefont
  {Cavagna}}, \bibinfo {author} {\bibfnamefont {L.}~\bibnamefont
  {Del~Castello}}, \bibinfo {author} {\bibfnamefont {I.}~\bibnamefont
  {Giardina}}, \bibinfo {author} {\bibfnamefont {T.}~\bibnamefont {Grigera}},
  \bibinfo {author} {\bibfnamefont {A.}~\bibnamefont {Jelic}}, \bibinfo
  {author} {\bibfnamefont {S.}~\bibnamefont {Melillo}}, \bibinfo {author}
  {\bibfnamefont {T.}~\bibnamefont {Mora}}, \bibinfo {author} {\bibfnamefont
  {L.}~\bibnamefont {Parisi}}, \bibinfo {author} {\bibfnamefont
  {E.}~\bibnamefont {Silvestri}}, \bibinfo {author} {\bibfnamefont
  {M.}~\bibnamefont {Viale}}, \ and\ \bibinfo {author} {\bibfnamefont {A.~M.}\
  \bibnamefont {Walczak}},\ }\href {\doibase 10.1007/s10955-014-1119-3}
  {\bibfield  {journal} {\bibinfo  {journal} {Journal of Statistical Physics}\
  }\textbf {\bibinfo {volume} {158}},\ \bibinfo {pages} {601} (\bibinfo {year}
  {2015})}\BibitemShut {NoStop}%
\bibitem [{\citenamefont {Young}\ \emph {et~al.}(2013)\citenamefont {Young},
  \citenamefont {Scardovi}, \citenamefont {Cavagna}, \citenamefont {Giardina},\
  and\ \citenamefont {Leonard}}]{Leonard2013}%
  \BibitemOpen
  \bibfield  {author} {\bibinfo {author} {\bibfnamefont {G.~F.}\ \bibnamefont
  {Young}}, \bibinfo {author} {\bibfnamefont {L.}~\bibnamefont {Scardovi}},
  \bibinfo {author} {\bibfnamefont {A.}~\bibnamefont {Cavagna}}, \bibinfo
  {author} {\bibfnamefont {I.}~\bibnamefont {Giardina}}, \ and\ \bibinfo
  {author} {\bibfnamefont {N.~E.}\ \bibnamefont {Leonard}},\ }\href {\doibase
  10.1371/journal.pcbi.1002894} {\bibfield  {journal} {\bibinfo  {journal}
  {PLoS Comput. Biol.}\ }\textbf {\bibinfo {volume} {9}},\ \bibinfo {pages} {1}
  (\bibinfo {year} {2013})}\BibitemShut {NoStop}%
\bibitem [{\citenamefont {Ballerini}\ \emph {et~al.}(2008)\citenamefont
  {Ballerini}, \citenamefont {Cabibbo}, \citenamefont {Candelier},
  \citenamefont {Cavagna}, \citenamefont {Cisbani}, \citenamefont {Giardina},
  \citenamefont {Lecomte}, \citenamefont {Orlandi}, \citenamefont {Parisi},
  \citenamefont {Procaccini}, \citenamefont {Viale},\ and\ \citenamefont
  {Zdravkovic}}]{Ballerini08}%
  \BibitemOpen
  \bibfield  {author} {\bibinfo {author} {\bibfnamefont {M.}~\bibnamefont
  {Ballerini}}, \bibinfo {author} {\bibfnamefont {N.}~\bibnamefont {Cabibbo}},
  \bibinfo {author} {\bibfnamefont {R.}~\bibnamefont {Candelier}}, \bibinfo
  {author} {\bibfnamefont {A.}~\bibnamefont {Cavagna}}, \bibinfo {author}
  {\bibfnamefont {E.}~\bibnamefont {Cisbani}}, \bibinfo {author} {\bibfnamefont
  {I.}~\bibnamefont {Giardina}}, \bibinfo {author} {\bibfnamefont
  {V.}~\bibnamefont {Lecomte}}, \bibinfo {author} {\bibfnamefont
  {A.}~\bibnamefont {Orlandi}}, \bibinfo {author} {\bibfnamefont
  {G.}~\bibnamefont {Parisi}}, \bibinfo {author} {\bibfnamefont
  {A.}~\bibnamefont {Procaccini}}, \bibinfo {author} {\bibfnamefont
  {M.}~\bibnamefont {Viale}}, \ and\ \bibinfo {author} {\bibfnamefont
  {V.}~\bibnamefont {Zdravkovic}},\ }\href {\doibase 10.1073/pnas.0711437105}
  {\bibfield  {journal} {\bibinfo  {journal} {Proc. Natl. Acad. Sci. U.S.A}\
  }\textbf {\bibinfo {volume} {105}},\ \bibinfo {pages} {1232} (\bibinfo {year}
  {2008})}\BibitemShut {NoStop}%
\bibitem [{\citenamefont {Ling}\ \emph {et~al.}(2019)\citenamefont {Ling},
  \citenamefont {E.}, \citenamefont {van~der Vaart~K.}, \citenamefont {T.},
  \citenamefont {A.},\ and\ \citenamefont {Ouellette}}]{Ouellette2019}%
  \BibitemOpen
  \bibfield  {author} {\bibinfo {author} {\bibfnamefont {H.}~\bibnamefont
  {Ling}}, \bibinfo {author} {\bibfnamefont {M.~G.}\ \bibnamefont {E.}},
  \bibinfo {author} {\bibnamefont {van~der Vaart~K.}}, \bibinfo {author}
  {\bibfnamefont {V.~R.}\ \bibnamefont {T.}}, \bibinfo {author} {\bibfnamefont
  {T.}~\bibnamefont {A.}}, \ and\ \bibinfo {author} {\bibfnamefont {N.~T.}\
  \bibnamefont {Ouellette}},\ }\href {\doibase 10.1098/rspb.2019.0865}
  {\bibfield  {journal} {\bibinfo  {journal} {R. Soc. B.}\ }\textbf {\bibinfo
  {volume} {286}} (\bibinfo {year} {2019}),\
  10.1098/rspb.2019.0865}\BibitemShut {NoStop}%
\bibitem [{\citenamefont {Li}\ and\ \citenamefont
  {Sayed}(2012)}]{Li_Sayed_2012}%
  \BibitemOpen
  \bibfield  {author} {\bibinfo {author} {\bibfnamefont {J.}~\bibnamefont
  {Li}}\ and\ \bibinfo {author} {\bibfnamefont {A.~H.}\ \bibnamefont {Sayed}},\
  }\href {\doibase 10.1186/1687-6180-2012-18} {\bibfield  {journal} {\bibinfo
  {journal} {EURASIP Journal on Advances in Signal Processing}\ }\textbf
  {\bibinfo {volume} {2012}},\ \bibinfo {pages} {18} (\bibinfo {year}
  {2012})}\BibitemShut {NoStop}%
\bibitem [{\citenamefont {Theraulaz}\ \emph {et~al.}(2002)\citenamefont
  {Theraulaz}, \citenamefont {Bonabeau}, \citenamefont {Nicolis}, \citenamefont
  {Sol{\'e}}, \citenamefont {Fourcassi{\'e}}, \citenamefont {Blanco},
  \citenamefont {Fournier}, \citenamefont {Joly}, \citenamefont
  {Fern{\'a}ndez}, \citenamefont {Grimal}, \citenamefont {Dalle},\ and\
  \citenamefont {Deneubourg}}]{Theraulaz2002}%
  \BibitemOpen
  \bibfield  {author} {\bibinfo {author} {\bibfnamefont {G.}~\bibnamefont
  {Theraulaz}}, \bibinfo {author} {\bibfnamefont {E.}~\bibnamefont {Bonabeau}},
  \bibinfo {author} {\bibfnamefont {S.~C.}\ \bibnamefont {Nicolis}}, \bibinfo
  {author} {\bibfnamefont {R.~V.}\ \bibnamefont {Sol{\'e}}}, \bibinfo {author}
  {\bibfnamefont {V.}~\bibnamefont {Fourcassi{\'e}}}, \bibinfo {author}
  {\bibfnamefont {S.}~\bibnamefont {Blanco}}, \bibinfo {author} {\bibfnamefont
  {R.}~\bibnamefont {Fournier}}, \bibinfo {author} {\bibfnamefont {J.-L.}\
  \bibnamefont {Joly}}, \bibinfo {author} {\bibfnamefont {P.}~\bibnamefont
  {Fern{\'a}ndez}}, \bibinfo {author} {\bibfnamefont {A.}~\bibnamefont
  {Grimal}}, \bibinfo {author} {\bibfnamefont {P.}~\bibnamefont {Dalle}}, \
  and\ \bibinfo {author} {\bibfnamefont {J.-L.}\ \bibnamefont {Deneubourg}},\
  }\href {\doibase 10.1073/pnas.152302199} {\bibfield  {journal} {\bibinfo
  {journal} {Proc. Natl. Acad. Sci. U.S.A}\ }\textbf {\bibinfo {volume} {99}},\
  \bibinfo {pages} {9645} (\bibinfo {year} {2002})}\BibitemShut {NoStop}%
\bibitem [{\citenamefont {Topaz}\ \emph {et~al.}(2012)\citenamefont {Topaz},
  \citenamefont {D'Orsogna}, \citenamefont {Edelstein-Keshet},\ and\
  \citenamefont {Bernoff}}]{Topaz2012}%
  \BibitemOpen
  \bibfield  {author} {\bibinfo {author} {\bibfnamefont {C.~M.}\ \bibnamefont
  {Topaz}}, \bibinfo {author} {\bibfnamefont {M.~R.}\ \bibnamefont
  {D'Orsogna}}, \bibinfo {author} {\bibfnamefont {L.}~\bibnamefont
  {Edelstein-Keshet}}, \ and\ \bibinfo {author} {\bibfnamefont {A.~J.}\
  \bibnamefont {Bernoff}},\ }\href {\doibase 10.1371/journal.pcbi.1002642}
  {\bibfield  {journal} {\bibinfo  {journal} {PLoS Comput. Biol.}\ }\textbf
  {\bibinfo {volume} {8}},\ \bibinfo {pages} {1} (\bibinfo {year}
  {2012})}\BibitemShut {NoStop}%
\bibitem [{\citenamefont {Polezhaev}\ \emph {et~al.}(2006)\citenamefont
  {Polezhaev}, \citenamefont {Pashkov}, \citenamefont {Lobanov},\ and\
  \citenamefont {Petrov}}]{Polezhaev}%
  \BibitemOpen
  \bibfield  {author} {\bibinfo {author} {\bibfnamefont {A.}~\bibnamefont
  {Polezhaev}}, \bibinfo {author} {\bibfnamefont {R.}~\bibnamefont {Pashkov}},
  \bibinfo {author} {\bibfnamefont {A.~I.}\ \bibnamefont {Lobanov}}, \ and\
  \bibinfo {author} {\bibfnamefont {I.~B.}\ \bibnamefont {Petrov}},\
  }\href@noop {} {\bibfield  {journal} {\bibinfo  {journal} {Int. J. Dev.
  Bio.}\ }\textbf {\bibinfo {volume} {50}},\ \bibinfo {pages} {309} (\bibinfo
  {year} {2006})}\BibitemShut {NoStop}%
\bibitem [{\citenamefont {Rio}\ and\ \citenamefont
  {Warren}(2014)}]{Rio_Warren_2014}%
  \BibitemOpen
  \bibfield  {author} {\bibinfo {author} {\bibfnamefont {K.}~\bibnamefont
  {Rio}}\ and\ \bibinfo {author} {\bibfnamefont {W.~H.}\ \bibnamefont
  {Warren}},\ }\href {\doibase https://doi.org/10.1016/j.trpro.2014.09.017}
  {\bibfield  {journal} {\bibinfo  {journal} {Transportation Research
  Procedia}\ }\textbf {\bibinfo {volume} {2}},\ \bibinfo {pages} {132 }
  (\bibinfo {year} {2014})},\ \bibinfo {note} {the Conference on Pedestrian and
  Evacuation Dynamics 2014 (PED 2014), 22-24 October 2014, Delft, The
  Netherlands}\BibitemShut {NoStop}%
\bibitem [{\citenamefont {Vicsek}\ and\ \citenamefont
  {Zafeiris}(2012)}]{Vicsek}%
  \BibitemOpen
  \bibfield  {author} {\bibinfo {author} {\bibfnamefont {T.}~\bibnamefont
  {Vicsek}}\ and\ \bibinfo {author} {\bibfnamefont {A.}~\bibnamefont
  {Zafeiris}},\ }\href@noop {} {\bibfield  {journal} {\bibinfo  {journal}
  {Phys. Rep.}\ }\textbf {\bibinfo {volume} {517}},\ \bibinfo {pages} {71}
  (\bibinfo {year} {2012})}\BibitemShut {NoStop}%
\bibitem [{\citenamefont {Marchetti}\ \emph {et~al.}(2013)\citenamefont
  {Marchetti}, \citenamefont {Joanny}, \citenamefont {Ramaswamy}, \citenamefont
  {Liverpool}, \citenamefont {Prost}, \citenamefont {Rao},\ and\ \citenamefont
  {Simha}}]{Marchetti}%
  \BibitemOpen
  \bibfield  {author} {\bibinfo {author} {\bibfnamefont {M.~C.}\ \bibnamefont
  {Marchetti}}, \bibinfo {author} {\bibfnamefont {J.~F.}\ \bibnamefont
  {Joanny}}, \bibinfo {author} {\bibfnamefont {S.}~\bibnamefont {Ramaswamy}},
  \bibinfo {author} {\bibfnamefont {T.~B.}\ \bibnamefont {Liverpool}}, \bibinfo
  {author} {\bibfnamefont {J.}~\bibnamefont {Prost}}, \bibinfo {author}
  {\bibfnamefont {M.}~\bibnamefont {Rao}}, \ and\ \bibinfo {author}
  {\bibfnamefont {R.~A.}\ \bibnamefont {Simha}},\ }\href@noop {} {\bibfield
  {journal} {\bibinfo  {journal} {Rev. Mod. Phys.}\ }\textbf {\bibinfo {volume}
  {85}},\ \bibinfo {pages} {1143} (\bibinfo {year} {2013})}\BibitemShut
  {NoStop}%
\bibitem [{\citenamefont {Aldana}\ \emph {et~al.}(2007)\citenamefont {Aldana},
  \citenamefont {Dossetti}, \citenamefont {Huepe}, \citenamefont {Kenkre},\
  and\ \citenamefont {Larralde}}]{Aldana}%
  \BibitemOpen
  \bibfield  {author} {\bibinfo {author} {\bibfnamefont {M.}~\bibnamefont
  {Aldana}}, \bibinfo {author} {\bibfnamefont {V.}~\bibnamefont {Dossetti}},
  \bibinfo {author} {\bibfnamefont {C.}~\bibnamefont {Huepe}}, \bibinfo
  {author} {\bibfnamefont {V.~M.}\ \bibnamefont {Kenkre}}, \ and\ \bibinfo
  {author} {\bibfnamefont {H.}~\bibnamefont {Larralde}},\ }\href@noop {}
  {\bibfield  {journal} {\bibinfo  {journal} {Phys. Rev. Letts.}\ }\textbf
  {\bibinfo {volume} {98}},\ \bibinfo {pages} {095702} (\bibinfo {year}
  {2007})}\BibitemShut {NoStop}%
\bibitem [{\citenamefont {Brambilla}\ \emph {et~al.}(2013)\citenamefont
  {Brambilla}, \citenamefont {Ferrante}, \citenamefont {Birattari},\ and\
  \citenamefont {Dorigo}}]{Brambilla2013}%
  \BibitemOpen
  \bibfield  {author} {\bibinfo {author} {\bibfnamefont {M.}~\bibnamefont
  {Brambilla}}, \bibinfo {author} {\bibfnamefont {E.}~\bibnamefont {Ferrante}},
  \bibinfo {author} {\bibfnamefont {M.}~\bibnamefont {Birattari}}, \ and\
  \bibinfo {author} {\bibfnamefont {M.}~\bibnamefont {Dorigo}},\ }\href
  {\doibase 10.1007/s11721-012-0075-2} {\bibfield  {journal} {\bibinfo
  {journal} {Swarm Intelligence}\ }\textbf {\bibinfo {volume} {7}},\ \bibinfo
  {pages} {1} (\bibinfo {year} {2013})}\BibitemShut {NoStop}%
\bibitem [{\citenamefont {Bayõndõr}(2016)}]{Bayindir16}%
  \BibitemOpen
  \bibfield  {author} {\bibinfo {author} {\bibfnamefont {L.}~\bibnamefont
  {Bayõndõr}},\ }\href {\doibase https://doi.org/10.1016/j.neucom.2015.05.116}
  {\bibfield  {journal} {\bibinfo  {journal} {Neurocomputing}\ }\textbf
  {\bibinfo {volume} {172}},\ \bibinfo {pages} {292 } (\bibinfo {year}
  {2016})}\BibitemShut {NoStop}%
\bibitem [{\citenamefont {Bandyopadhyay}(2005)}]{Bandyopadhyay}%
  \BibitemOpen
  \bibfield  {author} {\bibinfo {author} {\bibfnamefont {P.}~\bibnamefont
  {Bandyopadhyay}},\ }\href@noop {} {\bibfield  {journal} {\bibinfo  {journal}
  {IEEE Trans. Ocean. Engr.}\ }\textbf {\bibinfo {volume} {30}},\ \bibinfo
  {pages} {109} (\bibinfo {year} {2005})}\BibitemShut {NoStop}%
\bibitem [{\citenamefont {Wu}(2011)}]{Wu}%
  \BibitemOpen
  \bibfield  {author} {\bibinfo {author} {\bibfnamefont {W.}~\bibnamefont
  {Wu}},\ }\href@noop {} {\bibfield  {journal} {\bibinfo  {journal}
  {Automatica}\ }\textbf {\bibinfo {volume} {47}},\ \bibinfo {pages} {2044}
  (\bibinfo {year} {2011})}\BibitemShut {NoStop}%
\bibitem [{\citenamefont {Ramachandran}\ \emph {et~al.}(2018)\citenamefont
  {Ramachandran}, \citenamefont {Elamvazhuthi},\ and\ \citenamefont
  {Berman}}]{Ramachandran2018}%
  \BibitemOpen
  \bibfield  {author} {\bibinfo {author} {\bibfnamefont {R.~K.}\ \bibnamefont
  {Ramachandran}}, \bibinfo {author} {\bibfnamefont {K.}~\bibnamefont
  {Elamvazhuthi}}, \ and\ \bibinfo {author} {\bibfnamefont {S.}~\bibnamefont
  {Berman}},\ }\enquote {\bibinfo {title} {An optimal control approach to
  mapping gps-denied environments using a stochastic robotic swarm},}\ in\
  \href {\doibase 10.1007/978-3-319-51532-8_29} {\emph {\bibinfo {booktitle}
  {Robotics Research: Volume 1}}},\ \bibinfo {editor} {edited by\ \bibinfo
  {editor} {\bibfnamefont {A.}~\bibnamefont {Bicchi}}\ and\ \bibinfo {editor}
  {\bibfnamefont {W.}~\bibnamefont {Burgard}}}\ (\bibinfo  {publisher}
  {Springer International Publishing},\ \bibinfo {address} {Cham},\ \bibinfo
  {year} {2018})\ pp.\ \bibinfo {pages} {477--493}\BibitemShut {NoStop}%
\bibitem [{\citenamefont {Morgan}\ and\ \citenamefont
  {Schwartz}(2005)}]{MORGAN2005}%
  \BibitemOpen
  \bibfield  {author} {\bibinfo {author} {\bibfnamefont {D.~S.}\ \bibnamefont
  {Morgan}}\ and\ \bibinfo {author} {\bibfnamefont {I.~B.}\ \bibnamefont
  {Schwartz}},\ }\href {\doibase
  https://doi.org/10.1016/j.physleta.2005.03.074} {\bibfield  {journal}
  {\bibinfo  {journal} {Physics Letters A}\ }\textbf {\bibinfo {volume}
  {340}},\ \bibinfo {pages} {121 } (\bibinfo {year} {2005})}\BibitemShut
  {NoStop}%
\bibitem [{\citenamefont {Wiech}\ \emph {et~al.}(2018)\citenamefont {Wiech},
  \citenamefont {Eremeyev},\ and\ \citenamefont {Giorgio}}]{Wiech2018}%
  \BibitemOpen
  \bibfield  {author} {\bibinfo {author} {\bibfnamefont {J.}~\bibnamefont
  {Wiech}}, \bibinfo {author} {\bibfnamefont {V.~A.}\ \bibnamefont {Eremeyev}},
  \ and\ \bibinfo {author} {\bibfnamefont {I.}~\bibnamefont {Giorgio}},\ }\href
  {\doibase 10.1007/s00161-018-0664-4} {\bibfield  {journal} {\bibinfo
  {journal} {Continuum Mechanics and Thermodynamics}\ }\textbf {\bibinfo
  {volume} {30}},\ \bibinfo {pages} {1091} (\bibinfo {year}
  {2018})}\BibitemShut {NoStop}%
\bibitem [{\citenamefont {{Li}}\ \emph {et~al.}(2017)\citenamefont {{Li}},
  \citenamefont {{Feng}}, \citenamefont {{Ehrhard}}, \citenamefont {{Shen}},
  \citenamefont {{Cobos}}, \citenamefont {{Zhang}}, \citenamefont
  {{Elamvazhuthi}}, \citenamefont {{Berman}}, \citenamefont {{Haberland}},\
  and\ \citenamefont {{Bertozzi}}}]{Li17}%
  \BibitemOpen
  \bibfield  {author} {\bibinfo {author} {\bibfnamefont {H.}~\bibnamefont
  {{Li}}}, \bibinfo {author} {\bibfnamefont {C.}~\bibnamefont {{Feng}}},
  \bibinfo {author} {\bibfnamefont {H.}~\bibnamefont {{Ehrhard}}}, \bibinfo
  {author} {\bibfnamefont {Y.}~\bibnamefont {{Shen}}}, \bibinfo {author}
  {\bibfnamefont {B.}~\bibnamefont {{Cobos}}}, \bibinfo {author} {\bibfnamefont
  {F.}~\bibnamefont {{Zhang}}}, \bibinfo {author} {\bibfnamefont
  {K.}~\bibnamefont {{Elamvazhuthi}}}, \bibinfo {author} {\bibfnamefont
  {S.}~\bibnamefont {{Berman}}}, \bibinfo {author} {\bibfnamefont
  {M.}~\bibnamefont {{Haberland}}}, \ and\ \bibinfo {author} {\bibfnamefont
  {A.~L.}\ \bibnamefont {{Bertozzi}}},\ }in\ \href {\doibase
  10.1109/IROS.2017.8206299} {\emph {\bibinfo {booktitle} {2017 IEEE/RSJ
  International Conference on Intelligent Robots and Systems (IROS)}}}\
  (\bibinfo {year} {2017})\ pp.\ \bibinfo {pages} {4341--4347}\BibitemShut
  {NoStop}%
\bibitem [{\citenamefont {{Tanner}}\ \emph {et~al.}(2007)\citenamefont
  {{Tanner}}, \citenamefont {{Jadbabaie}},\ and\ \citenamefont
  {{Pappas}}}]{Tanner07}%
  \BibitemOpen
  \bibfield  {author} {\bibinfo {author} {\bibfnamefont {H.~G.}\ \bibnamefont
  {{Tanner}}}, \bibinfo {author} {\bibfnamefont {A.}~\bibnamefont
  {{Jadbabaie}}}, \ and\ \bibinfo {author} {\bibfnamefont {G.~J.}\ \bibnamefont
  {{Pappas}}},\ }\href {\doibase 10.1109/TAC.2007.895948} {\bibfield  {journal}
  {\bibinfo  {journal} {IEEE Transactions on Automatic Control}\ }\textbf
  {\bibinfo {volume} {52}},\ \bibinfo {pages} {863} (\bibinfo {year}
  {2007})}\BibitemShut {NoStop}%
\bibitem [{\citenamefont {{Gazi}}(2005)}]{Gazi05}%
  \BibitemOpen
  \bibfield  {author} {\bibinfo {author} {\bibfnamefont {V.}~\bibnamefont
  {{Gazi}}},\ }\href {\doibase 10.1109/TRO.2005.853487} {\bibfield  {journal}
  {\bibinfo  {journal} {IEEE Transactions on Robotics}\ }\textbf {\bibinfo
  {volume} {21}},\ \bibinfo {pages} {1208} (\bibinfo {year}
  {2005})}\BibitemShut {NoStop}%
\bibitem [{\citenamefont {{Jadbabaie}}\ \emph {et~al.}(2003)\citenamefont
  {{Jadbabaie}}, \citenamefont {{Jie Lin}},\ and\ \citenamefont
  {{Morse}}}]{Jadbabaie03}%
  \BibitemOpen
  \bibfield  {author} {\bibinfo {author} {\bibfnamefont {A.}~\bibnamefont
  {{Jadbabaie}}}, \bibinfo {author} {\bibnamefont {{Jie Lin}}}, \ and\ \bibinfo
  {author} {\bibfnamefont {A.~S.}\ \bibnamefont {{Morse}}},\ }\href {\doibase
  10.1109/TAC.2003.812781} {\bibfield  {journal} {\bibinfo  {journal} {IEEE
  Transactions on Automatic Control}\ }\textbf {\bibinfo {volume} {48}},\
  \bibinfo {pages} {988} (\bibinfo {year} {2003})}\BibitemShut {NoStop}%
\bibitem [{\citenamefont {Vir{\'{a}}gh}\ \emph {et~al.}(2014)\citenamefont
  {Vir{\'{a}}gh}, \citenamefont {V{\'{a}}s{\'{a}}rhelyi}, \citenamefont
  {Tarcai}, \citenamefont {Ször{\'{e}}nyi}, \citenamefont {Somorjai},
  \citenamefont {Nepusz},\ and\ \citenamefont {Vicsek}}]{Viraghn14}%
  \BibitemOpen
  \bibfield  {author} {\bibinfo {author} {\bibfnamefont {C.}~\bibnamefont
  {Vir{\'{a}}gh}}, \bibinfo {author} {\bibfnamefont {G.}~\bibnamefont
  {V{\'{a}}s{\'{a}}rhelyi}}, \bibinfo {author} {\bibfnamefont {N.}~\bibnamefont
  {Tarcai}}, \bibinfo {author} {\bibfnamefont {T.}~\bibnamefont
  {Ször{\'{e}}nyi}}, \bibinfo {author} {\bibfnamefont {G.}~\bibnamefont
  {Somorjai}}, \bibinfo {author} {\bibfnamefont {T.}~\bibnamefont {Nepusz}}, \
  and\ \bibinfo {author} {\bibfnamefont {T.}~\bibnamefont {Vicsek}},\ }\href
  {\doibase 10.1088/1748-3182/9/2/025012} {\bibfield  {journal} {\bibinfo
  {journal} {Bioinspiration {\&} Biomimetics}\ }\textbf {\bibinfo {volume}
  {9}},\ \bibinfo {pages} {025012} (\bibinfo {year} {2014})}\BibitemShut
  {NoStop}%
\bibitem [{\citenamefont {{Desai}}\ \emph {et~al.}(2001)\citenamefont
  {{Desai}}, \citenamefont {{Ostrowski}},\ and\ \citenamefont
  {{Kumar}}}]{Desai01}%
  \BibitemOpen
  \bibfield  {author} {\bibinfo {author} {\bibfnamefont {J.~P.}\ \bibnamefont
  {{Desai}}}, \bibinfo {author} {\bibfnamefont {J.~P.}\ \bibnamefont
  {{Ostrowski}}}, \ and\ \bibinfo {author} {\bibfnamefont {V.}~\bibnamefont
  {{Kumar}}},\ }in\ \href@noop {} {\emph {\bibinfo {booktitle} {IEEE
  Transactions on Robotics and Automation}}},\ Vol.\ \bibinfo {volume} {17(6)}\
  (\bibinfo {year} {2001})\ pp.\ \bibinfo {pages} {905--908}\BibitemShut
  {NoStop}%
\bibitem [{\citenamefont {Giuggioli}\ \emph {et~al.}(2015)\citenamefont
  {Giuggioli}, \citenamefont {McKetterick},\ and\ \citenamefont
  {Holderied}}]{Luca_bats}%
  \BibitemOpen
  \bibfield  {author} {\bibinfo {author} {\bibfnamefont {L.}~\bibnamefont
  {Giuggioli}}, \bibinfo {author} {\bibfnamefont {T.}~\bibnamefont
  {McKetterick}}, \ and\ \bibinfo {author} {\bibfnamefont {M.}~\bibnamefont
  {Holderied}},\ }\href@noop {} {\bibfield  {journal} {\bibinfo  {journal}
  {PLoS Comput. Biol.}\ }\textbf {\bibinfo {volume} {11}} (\bibinfo {year}
  {2015})}\BibitemShut {NoStop}%
\bibitem [{\citenamefont {Nagy}\ \emph {et~al.}(2010)\citenamefont {Nagy},
  \citenamefont {Akos}, \citenamefont {Biro},\ and\ \citenamefont
  {Vicsek}}]{Nagy_pigeons}%
  \BibitemOpen
  \bibfield  {author} {\bibinfo {author} {\bibfnamefont {N.}~\bibnamefont
  {Nagy}}, \bibinfo {author} {\bibfnamefont {Z.}~\bibnamefont {Akos}}, \bibinfo
  {author} {\bibfnamefont {D.}~\bibnamefont {Biro}}, \ and\ \bibinfo {author}
  {\bibfnamefont {T.}~\bibnamefont {Vicsek}},\ }\href@noop {} {\bibfield
  {journal} {\bibinfo  {journal} {Nature}\ }\textbf {\bibinfo {volume} {464}},\
  \bibinfo {pages} {890} (\bibinfo {year} {2010})}\BibitemShut {NoStop}%
\bibitem [{\citenamefont {Jiang}\ \emph {et~al.}(2017)\citenamefont {Jiang},
  \citenamefont {Giuggioli}, \citenamefont {Perna},\ and\ \citenamefont {et.
  al.}}]{JL_fish}%
  \BibitemOpen
  \bibfield  {author} {\bibinfo {author} {\bibfnamefont {L.}~\bibnamefont
  {Jiang}}, \bibinfo {author} {\bibfnamefont {L.}~\bibnamefont {Giuggioli}},
  \bibinfo {author} {\bibfnamefont {A.}~\bibnamefont {Perna}}, \ and\ \bibinfo
  {author} {\bibnamefont {et. al.}},\ }\href@noop {} {\bibfield  {journal}
  {\bibinfo  {journal} {PLoS Comput. Biol.}\ }\textbf {\bibinfo {volume}
  {13}},\ \bibinfo {pages} {e1005822} (\bibinfo {year} {2017})}\BibitemShut
  {NoStop}%
\bibitem [{\citenamefont {Fehrenbach}\ \emph {et~al.}(2014)\citenamefont
  {Fehrenbach}, \citenamefont {Narski}, \citenamefont {Hua}, \citenamefont
  {Lemercier}, \citenamefont {Jelic}, \citenamefont {Appert-Rolland},
  \citenamefont {Donikian}, \citenamefont {Pettré},\ and\ \citenamefont
  {Degond}}]{JF_people}%
  \BibitemOpen
  \bibfield  {author} {\bibinfo {author} {\bibfnamefont {J.}~\bibnamefont
  {Fehrenbach}}, \bibinfo {author} {\bibfnamefont {J.}~\bibnamefont {Narski}},
  \bibinfo {author} {\bibfnamefont {J.}~\bibnamefont {Hua}}, \bibinfo {author}
  {\bibfnamefont {S.}~\bibnamefont {Lemercier}}, \bibinfo {author}
  {\bibfnamefont {A.}~\bibnamefont {Jelic}}, \bibinfo {author} {\bibfnamefont
  {C.}~\bibnamefont {Appert-Rolland}}, \bibinfo {author} {\bibfnamefont
  {S.}~\bibnamefont {Donikian}}, \bibinfo {author} {\bibfnamefont
  {J.}~\bibnamefont {Pettré}}, \ and\ \bibinfo {author} {\bibfnamefont
  {P.}~\bibnamefont {Degond}},\ }\href {\doibase 10.3934/nhm.2015.10.579} {\
  (\bibinfo {year} {2014}),\ 10.3934/nhm.2015.10.579},\ \Eprint
  {http://arxiv.org/abs/arXiv:1412.7537} {arXiv:1412.7537} \BibitemShut
  {NoStop}%
\bibitem [{\citenamefont {Komareji}\ \emph {et~al.}(2018)\citenamefont
  {Komareji}, \citenamefont {Shang},\ and\ \citenamefont
  {Bouffanais}}]{Komareji2018}%
  \BibitemOpen
  \bibfield  {author} {\bibinfo {author} {\bibfnamefont {M.}~\bibnamefont
  {Komareji}}, \bibinfo {author} {\bibfnamefont {Y.}~\bibnamefont {Shang}}, \
  and\ \bibinfo {author} {\bibfnamefont {R.}~\bibnamefont {Bouffanais}},\
  }\href {\doibase 10.1007/s11071-018-4259-1} {\bibfield  {journal} {\bibinfo
  {journal} {Nonlinear Dynamics}\ }\textbf {\bibinfo {volume} {93}},\ \bibinfo
  {pages} {1287} (\bibinfo {year} {2018})}\BibitemShut {NoStop}%
\bibitem [{\citenamefont {{Oliveira}}\ \emph {et~al.}(2015)\citenamefont
  {{Oliveira}}, \citenamefont {{Almeida}},\ and\ \citenamefont
  {{Lima}}}]{Oliveira15}%
  \BibitemOpen
  \bibfield  {author} {\bibinfo {author} {\bibfnamefont {L.}~\bibnamefont
  {{Oliveira}}}, \bibinfo {author} {\bibfnamefont {L.}~\bibnamefont
  {{Almeida}}}, \ and\ \bibinfo {author} {\bibfnamefont {P.}~\bibnamefont
  {{Lima}}},\ }in\ \href {\doibase 10.1109/WFCS.2015.7160566} {\emph {\bibinfo
  {booktitle} {2015 IEEE World Conference on Factory Communication Systems
  (WFCS)}}}\ (\bibinfo {year} {2015})\ pp.\ \bibinfo {pages} {1--8}\BibitemShut
  {NoStop}%
\bibitem [{\citenamefont {y~Teran-Romero}\ \emph {et~al.}(2012)\citenamefont
  {y~Teran-Romero}, \citenamefont {Forgoston},\ and\ \citenamefont
  {Schwartz}}]{Romero2012}%
  \BibitemOpen
  \bibfield  {author} {\bibinfo {author} {\bibfnamefont {L.~M.}\ \bibnamefont
  {y~Teran-Romero}}, \bibinfo {author} {\bibfnamefont {E.}~\bibnamefont
  {Forgoston}}, \ and\ \bibinfo {author} {\bibfnamefont {I.~B.}\ \bibnamefont
  {Schwartz}},\ }\href {\doibase 10.1109/TRO.2012.2198511} {\bibfield
  {journal} {\bibinfo  {journal} {IEEE Transactions on Robotics}\ }\textbf
  {\bibinfo {volume} {28}},\ \bibinfo {pages} {1034} (\bibinfo {year}
  {2012})}\BibitemShut {NoStop}%
\bibitem [{\citenamefont {Szwaykowska}\ \emph {et~al.}(2016)\citenamefont
  {Szwaykowska}, \citenamefont {Schwartz}, \citenamefont {Mier-y Teran~Romero},
  \citenamefont {Heckman}, \citenamefont {Mox},\ and\ \citenamefont
  {Hsieh}}]{Szwaykowska2016}%
  \BibitemOpen
  \bibfield  {author} {\bibinfo {author} {\bibfnamefont {K.}~\bibnamefont
  {Szwaykowska}}, \bibinfo {author} {\bibfnamefont {I.~B.}\ \bibnamefont
  {Schwartz}}, \bibinfo {author} {\bibfnamefont {L.}~\bibnamefont {Mier-y
  Teran~Romero}}, \bibinfo {author} {\bibfnamefont {C.~R.}\ \bibnamefont
  {Heckman}}, \bibinfo {author} {\bibfnamefont {D.}~\bibnamefont {Mox}}, \ and\
  \bibinfo {author} {\bibfnamefont {M.~A.}\ \bibnamefont {Hsieh}},\ }\href
  {\doibase 10.1103/PhysRevE.93.032307} {\bibfield  {journal} {\bibinfo
  {journal} {Phys. Rev. E}\ }\textbf {\bibinfo {volume} {93}},\ \bibinfo
  {pages} {032307} (\bibinfo {year} {2016})}\BibitemShut {NoStop}%
\bibitem [{\citenamefont {Edwards}\ \emph {et~al.}()\citenamefont {Edwards},
  \citenamefont {deZonia}, \citenamefont {Hsieh}, \citenamefont {Hindes},
  \citenamefont {Triandof},\ and\ \citenamefont {Schwartz}}]{Edwards2019}%
  \BibitemOpen
  \bibfield  {author} {\bibinfo {author} {\bibfnamefont {V.}~\bibnamefont
  {Edwards}}, \bibinfo {author} {\bibfnamefont {P.}~\bibnamefont {deZonia}},
  \bibinfo {author} {\bibfnamefont {M.~A.}\ \bibnamefont {Hsieh}}, \bibinfo
  {author} {\bibfnamefont {J.}~\bibnamefont {Hindes}}, \bibinfo {author}
  {\bibfnamefont {I.}~\bibnamefont {Triandof}}, \ and\ \bibinfo {author}
  {\bibfnamefont {I.~B.}\ \bibnamefont {Schwartz}},\ }\href@noop {} {\bibinfo
  {journal} {{\it Delay-Induced Swarm Pattern Bifurcations in Mixed-Reality
  Experiments}, Chaos [under review]}\ }\BibitemShut {NoStop}%
\bibitem [{\citenamefont {Wells}\ \emph {et~al.}(2015)\citenamefont {Wells},
  \citenamefont {Kath},\ and\ \citenamefont {Motter}}]{Wells15}%
  \BibitemOpen
\bibfield  {journal} {  }\bibfield  {author} {\bibinfo {author} {\bibfnamefont
  {D.~K.}\ \bibnamefont {Wells}}, \bibinfo {author} {\bibfnamefont {W.~L.}\
  \bibnamefont {Kath}}, \ and\ \bibinfo {author} {\bibfnamefont {A.~E.}\
  \bibnamefont {Motter}},\ }\href {\doibase 10.1103/PhysRevX.5.031036}
  {\bibfield  {journal} {\bibinfo  {journal} {Phys. Rev. X}\ }\textbf {\bibinfo
  {volume} {5}},\ \bibinfo {pages} {031036} (\bibinfo {year}
  {2015})}\BibitemShut {NoStop}%
\bibitem [{\citenamefont {Hartle}\ and\ \citenamefont
  {Wackerbauer}(2017)}]{Hartle17}%
  \BibitemOpen
  \bibfield  {author} {\bibinfo {author} {\bibfnamefont {H.}~\bibnamefont
  {Hartle}}\ and\ \bibinfo {author} {\bibfnamefont {R.}~\bibnamefont
  {Wackerbauer}},\ }\href {\doibase 10.1103/PhysRevE.96.032223} {\bibfield
  {journal} {\bibinfo  {journal} {Phys. Rev. E}\ }\textbf {\bibinfo {volume}
  {96}},\ \bibinfo {pages} {032223} (\bibinfo {year} {2017})}\BibitemShut
  {NoStop}%
\bibitem [{\citenamefont {Ansmann}\ \emph {et~al.}(2016)\citenamefont
  {Ansmann}, \citenamefont {Lehnertz},\ and\ \citenamefont
  {Feudel}}]{Ansmann16}%
  \BibitemOpen
  \bibfield  {author} {\bibinfo {author} {\bibfnamefont {G.}~\bibnamefont
  {Ansmann}}, \bibinfo {author} {\bibfnamefont {K.}~\bibnamefont {Lehnertz}}, \
  and\ \bibinfo {author} {\bibfnamefont {U.}~\bibnamefont {Feudel}},\ }\href
  {\doibase 10.1103/PhysRevX.6.011030} {\bibfield  {journal} {\bibinfo
  {journal} {Phys. Rev. X}\ }\textbf {\bibinfo {volume} {6}},\ \bibinfo {pages}
  {011030} (\bibinfo {year} {2016})}\BibitemShut {NoStop}%
\bibitem [{\citenamefont {Hindes}\ and\ \citenamefont
  {Schwartz}(2018)}]{Hindes18}%
  \BibitemOpen
  \bibfield  {author} {\bibinfo {author} {\bibfnamefont {J.}~\bibnamefont
  {Hindes}}\ and\ \bibinfo {author} {\bibfnamefont {I.~B.}\ \bibnamefont
  {Schwartz}},\ }\href {\doibase 10.1063/1.5041377} {\bibfield  {journal}
  {\bibinfo  {journal} {Chaos}\ }\textbf {\bibinfo {volume} {28}},\ \bibinfo
  {pages} {071106} (\bibinfo {year} {2018})}\BibitemShut {NoStop}%
\bibitem [{\citenamefont {{Szwaykowska}}\ \emph {et~al.}(2018)\citenamefont
  {{Szwaykowska}}, \citenamefont {{Schwartz}},\ and\ \citenamefont
  {{Carr}}}]{Szwaykowska2018}%
  \BibitemOpen
  \bibfield  {author} {\bibinfo {author} {\bibfnamefont {K.}~\bibnamefont
  {{Szwaykowska}}}, \bibinfo {author} {\bibfnamefont {I.~B.}\ \bibnamefont
  {{Schwartz}}}, \ and\ \bibinfo {author} {\bibfnamefont {T.~W.}\ \bibnamefont
  {{Carr}}},\ }in\ \href {\doibase 10.1109/ISMA.2018.8330138} {\emph {\bibinfo
  {booktitle} {11th International Symposium on Mechatronics and its
  Applications (ISMA)}}}\ (\bibinfo {year} {2018})\ pp.\ \bibinfo {pages}
  {1--6}\BibitemShut {NoStop}%
\bibitem [{\citenamefont {Kularatne}\ \emph {et~al.}(2019)\citenamefont
  {Kularatne}, \citenamefont {Forgoston},\ and\ \citenamefont
  {Hsieh}}]{Kularatne19}%
  \BibitemOpen
  \bibfield  {author} {\bibinfo {author} {\bibfnamefont {D.}~\bibnamefont
  {Kularatne}}, \bibinfo {author} {\bibfnamefont {E.}~\bibnamefont
  {Forgoston}}, \ and\ \bibinfo {author} {\bibfnamefont {M.~A.}\ \bibnamefont
  {Hsieh}},\ }\href {\doibase 10.1063/1.5090113} {\bibfield  {journal}
  {\bibinfo  {journal} {Chaos}\ }\textbf {\bibinfo {volume} {29}},\ \bibinfo
  {pages} {053128} (\bibinfo {year} {2019})}\BibitemShut {NoStop}%
\bibitem [{\citenamefont {Levine}\ \emph {et~al.}(2000)\citenamefont {Levine},
  \citenamefont {Rappel},\ and\ \citenamefont {Cohen}}]{Levine}%
  \BibitemOpen
  \bibfield  {author} {\bibinfo {author} {\bibfnamefont {H.}~\bibnamefont
  {Levine}}, \bibinfo {author} {\bibfnamefont {W.~J.}\ \bibnamefont {Rappel}},
  \ and\ \bibinfo {author} {\bibfnamefont {I.}~\bibnamefont {Cohen}},\
  }\href@noop {} {\bibfield  {journal} {\bibinfo  {journal} {Phys. Rev. E}\
  }\textbf {\bibinfo {volume} {63}},\ \bibinfo {pages} {017101} (\bibinfo
  {year} {2000})}\BibitemShut {NoStop}%
\bibitem [{\citenamefont {Erdmann}\ \emph {et~al.}(2005)\citenamefont
  {Erdmann}, \citenamefont {Ebeling},\ and\ \citenamefont
  {Mikhailov}}]{Erdmann}%
  \BibitemOpen
  \bibfield  {author} {\bibinfo {author} {\bibfnamefont {U.}~\bibnamefont
  {Erdmann}}, \bibinfo {author} {\bibfnamefont {W.}~\bibnamefont {Ebeling}}, \
  and\ \bibinfo {author} {\bibfnamefont {A.~S.}\ \bibnamefont {Mikhailov}},\
  }\href@noop {} {\bibfield  {journal} {\bibinfo  {journal} {Phys. Rev. E}\
  }\textbf {\bibinfo {volume} {71}},\ \bibinfo {pages} {051904} (\bibinfo
  {year} {2005})}\BibitemShut {NoStop}%
\bibitem [{\citenamefont {D'Orsogna}\ \emph {et~al.}(2006)\citenamefont
  {D'Orsogna}, \citenamefont {Chuang}, \citenamefont {Bertozzi},\ and\
  \citenamefont {Chayes}}]{DOrsagna}%
  \BibitemOpen
  \bibfield  {author} {\bibinfo {author} {\bibfnamefont {M.~R.}\ \bibnamefont
  {D'Orsogna}}, \bibinfo {author} {\bibfnamefont {Y.~L.}\ \bibnamefont
  {Chuang}}, \bibinfo {author} {\bibfnamefont {A.~L.}\ \bibnamefont
  {Bertozzi}}, \ and\ \bibinfo {author} {\bibfnamefont {L.~S.}\ \bibnamefont
  {Chayes}},\ }\href@noop {} {\bibfield  {journal} {\bibinfo  {journal} {Phys.
  Rev. Lett.}\ }\textbf {\bibinfo {volume} {96}},\ \bibinfo {pages} {104302}
  (\bibinfo {year} {2006})}\BibitemShut {NoStop}%
\bibitem [{\citenamefont {Forgoston}\ and\ \citenamefont
  {Schwartz}(2008)}]{F1}%
  \BibitemOpen
  \bibfield  {author} {\bibinfo {author} {\bibfnamefont {E.}~\bibnamefont
  {Forgoston}}\ and\ \bibinfo {author} {\bibfnamefont {I.~B.}\ \bibnamefont
  {Schwartz}},\ }\href@noop {} {\bibfield  {journal} {\bibinfo  {journal}
  {Phys. Rev. E}\ }\textbf {\bibinfo {volume} {77}},\ \bibinfo {pages}
  {035203(R)} (\bibinfo {year} {2008})}\BibitemShut {NoStop}%
\bibitem [{\citenamefont {Chuang}\ \emph {et~al.}(2007)\citenamefont {Chuang},
  \citenamefont {D'Orsogna}, \citenamefont {Marthaler}, \citenamefont
  {Bertozzi},\ and\ \citenamefont {Chayes}}]{Chuang}%
  \BibitemOpen
  \bibfield  {author} {\bibinfo {author} {\bibfnamefont {Y.}~\bibnamefont
  {Chuang}}, \bibinfo {author} {\bibfnamefont {M.}~\bibnamefont {D'Orsogna}},
  \bibinfo {author} {\bibfnamefont {D.}~\bibnamefont {Marthaler}}, \bibinfo
  {author} {\bibfnamefont {A.}~\bibnamefont {Bertozzi}}, \ and\ \bibinfo
  {author} {\bibfnamefont {L.}~\bibnamefont {Chayes}},\ }\href@noop {}
  {\bibfield  {journal} {\bibinfo  {journal} {Physica D}\ }\textbf {\bibinfo
  {volume} {232}},\ \bibinfo {pages} {33} (\bibinfo {year} {2007})}\BibitemShut
  {NoStop}%
\bibitem [{\citenamefont {Bernoff}\ and\ \citenamefont
  {Topaz}(2011)}]{Bernoff}%
  \BibitemOpen
  \bibfield  {author} {\bibinfo {author} {\bibfnamefont {A.}~\bibnamefont
  {Bernoff}}\ and\ \bibinfo {author} {\bibfnamefont {C.}~\bibnamefont
  {Topaz}},\ }\href@noop {} {\bibfield  {journal} {\bibinfo  {journal} {SIAM
  Journal of Applied Dynamical Systems}\ }\textbf {\bibinfo {volume} {10}},\
  \bibinfo {pages} {212} (\bibinfo {year} {2011})}\BibitemShut {NoStop}%
\bibitem [{\citenamefont {Hindes}\ \emph {et~al.}(2016)\citenamefont {Hindes},
  \citenamefont {Szwaykowska},\ and\ \citenamefont {Schwartz}}]{J1}%
  \BibitemOpen
  \bibfield  {author} {\bibinfo {author} {\bibfnamefont {J.}~\bibnamefont
  {Hindes}}, \bibinfo {author} {\bibfnamefont {K.}~\bibnamefont {Szwaykowska}},
  \ and\ \bibinfo {author} {\bibfnamefont {I.~B.}\ \bibnamefont {Schwartz}},\
  }\href@noop {} {\bibfield  {journal} {\bibinfo  {journal} {Phys. Rev. E}\
  }\textbf {\bibinfo {volume} {94}},\ \bibinfo {pages} {032306} (\bibinfo
  {year} {2016})}\BibitemShut {NoStop}%
\bibitem [{\citenamefont {{Lindley}}\ \emph {et~al.}(2013)\citenamefont
  {{Lindley}}, \citenamefont {{Mier-y-Teran-Romero}},\ and\ \citenamefont
  {{Schwartz}}}]{Lindley2013}%
  \BibitemOpen
  \bibfield  {author} {\bibinfo {author} {\bibfnamefont {B.}~\bibnamefont
  {{Lindley}}}, \bibinfo {author} {\bibfnamefont {L.}~\bibnamefont
  {{Mier-y-Teran-Romero}}}, \ and\ \bibinfo {author} {\bibfnamefont {I.~B.}\
  \bibnamefont {{Schwartz}}},\ }in\ \href {\doibase 10.1109/ACC.2013.6580546}
  {\emph {\bibinfo {booktitle} {2013 American Control Conference}}}\ (\bibinfo
  {year} {2013})\ pp.\ \bibinfo {pages} {4587--4591}\BibitemShut {NoStop}%
\bibitem [{\citenamefont {y~Teran-Romero}\ and\ \citenamefont
  {Schwartz}(2014)}]{Romero2014}%
  \BibitemOpen
  \bibfield  {author} {\bibinfo {author} {\bibfnamefont {L.~M.}\ \bibnamefont
  {y~Teran-Romero}}\ and\ \bibinfo {author} {\bibfnamefont {I.~B.}\
  \bibnamefont {Schwartz}},\ }\href {\doibase 10.1209/0295-5075/105/20002}
  {\bibfield  {journal} {\bibinfo  {journal} {{EPL}}\ }\textbf {\bibinfo
  {volume} {105}},\ \bibinfo {pages} {20002} (\bibinfo {year}
  {2014})}\BibitemShut {NoStop}%
\bibitem [{Note2()}]{Note2}%
  \BibitemOpen
  \bibinfo {note} {In this work the $N\!\gg\!1$ approximation implies replacing
  the restricted sum in Eq.(1), over all but one of the agents, to a sum over
  all agents.}\BibitemShut {Stop}%
\bibitem [{\citenamefont {Kuznetsov}(2004)}]{Kuznetsov1}%
  \BibitemOpen
  \bibfield  {author} {\bibinfo {author} {\bibfnamefont {Y.~A.}\ \bibnamefont
  {Kuznetsov}},\ }\href@noop {} {\emph {\bibinfo {title} {Elements of Applied
  Bifurcation Theory}}}\ (\bibinfo  {publisher} {Springer, Berlin},\ \bibinfo
  {year} {2004})\BibitemShut {NoStop}%
\bibitem [{Note1()}]{Note1}%
  \BibitemOpen
  \bibinfo {note} {Because the stability analysis is performed in rotating
  frames of reference, technically the Hopf bifurcations are torus
  bifurcations, and the saddle-node bifurcations are
  saddle-nodes-of-periodic-orbits}\BibitemShut {NoStop}%
\bibitem [{\citenamefont {Hindes}\ and\ \citenamefont {Myers}(2015)}]{J2}%
  \BibitemOpen
  \bibfield  {author} {\bibinfo {author} {\bibfnamefont {J.}~\bibnamefont
  {Hindes}}\ and\ \bibinfo {author} {\bibfnamefont {C.~R.}\ \bibnamefont
  {Myers}},\ }\href@noop {} {\bibfield  {journal} {\bibinfo  {journal} {Chaos}\
  }\textbf {\bibinfo {volume} {25}},\ \bibinfo {pages} {073119} (\bibinfo
  {year} {2015})}\BibitemShut {NoStop}%
\end{thebibliography}%
\end{document}